# A Review of Data-driven Approaches for Malicious Website Detection

Zeyuan Hu
*Intelligent Product Department*
*China Mobile (Zhejiang) Research*
*& Innovation Institute*
Hangzhou, China
huzeyuan@zj.chinamobile.com

Ziang Yuan
*Intelligent Product Department*
*China Mobile (Zhejiang) Research*
*& Innovation Institute*
Hangzhou, China
yuanziang2@zj.chinamobile.com

*Abstract*—The detection of malicious websites has become a critical issue in cybersecurity. Therefore, this paper offers a comprehensive review of data-driven methods for detecting malicious websites. Traditional approaches and their limitations are discussed, followed by an overview of data-driven approaches. The paper establishes the data-feature-model-extension pipeline and the latest research developments of data-driven approaches, including data preprocessing, feature extraction, model construction and technology extension. Specifically, this paper compares methods using deep learning models proposed in recent years. Furthermore, the paper follows the data-feature-model-extension pipeline to discuss the challenges together with some future directions of data-driven methods in malicious website detection.

*Keywords—malicious website detection, data-driven, deep learning, cybersecurity, pipeline*

## I. Introduction

### A. Background of Malicious Website Detection

With the increasing reliance on the internet for everyday tasks such as online shopping, banking, and socializing, the risk of encountering malicious websites has also increased. Specifically, malicious websites refer to web pages that are designed to harm or exploit users who visit them. This can include web pages that host malware, phishing pages that steal sensitive information, or pages that use social engineering tactics to trick users into performing harmful actions.

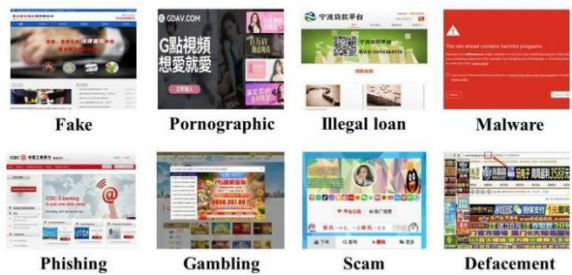

Fig. 1. Common malicious websites in China.

The detection of malicious websites is essential for protecting users from cyber threats. Malicious websites can cause significant damage to both individuals and organizations, including financial loss, data breaches, and reputational damage. In Figure 2, we present the trends in the daily total number of websites and malicious websites in the mobile communication traffic of Zhejiang Province, China, from January 2022 to April 2023.

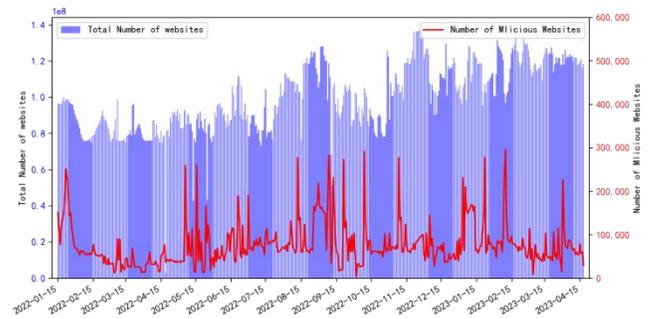

Fig. 2. The number of total websites and malicious websites in a province.

By detecting and blocking malicious websites, it is possible to mitigate the risks associated with these threats and ensure a safer online environment for users and enterprises. However, with the abundance of internet resources and the diversity of content in the online world, quickly and efficiently detecting malicious websites from a vast number of websites and accurately classifying them based on complex website content remains a major challenge in practical engineering applications.

### B. Traditional Detection Techniques

To ensure the security and privacy of internet users, various techniques have been developed to detect and block malicious websites. Two traditional approaches to detect malicious websites are blacklist-based and heuristic-based.

The blacklist-based methods maintain a list of known malicious websites, commonly called a blacklist, to compare the URL of new websites. A database search is done when a new website is accessed. If the URL of the website is listed on the blacklist, a warning signal will be generated. Due to their simplicity and effectiveness, blacklist-based approaches are used by many anti-virus software programs [1] and internet browsers. However, this technique has some limitations. First,

This work is funded by National Key Research and Development Project (Grant No: 2022YFB2703104).

this approach is exhaustive list maintenance [2] since the blacklist needs to be continuously updated to reflect the ever-evolving landscape of malicious websites. Second, attackers can evade detection by using new domains, subdomains, or IP addresses. In addition, the blacklist-based approach is susceptible to false negatives when the malicious websites are not on the list.

The heuristic-based methods are extensions of blacklist-based methods, which use several techniques to identify malicious websites, such as signature-based, content-based, or behavior-based analysis [3]. When compared to blacklist methods these methods offer better generalization capabilities since it is possible to detect new cyber threats on new websites. However, these techniques have the drawback that they were created specifically for a limited number of prevalent threats and cannot be generalized to all types of newer malicious website.

### C. Introduction to Data-driven Approaches

The emergence of data-driven techniques, represented by machine learning (ML) algorithms, serves a promising approach to detect malicious websites. These techniques can learn from large amounts of data, enabling them to detect previously unknown malicious websites and making them more effective than traditional techniques such as blacklist-based approaches and heuristic-based approaches. Additionally, data-driven approaches can reduce the number of false positives, leading to a more efficient use of resources. This has led to a growing attention in and recent research focus on the use of data-driven approaches in malicious website detection.

## II. DATA-DRIVEN APPROACHES FOR MALICIOUS WEBSITE DETECTION

Data-driven methods involve the use of machine learning, deep learning, statistical models, and data mining techniques to extract patterns and relationships from the data [4]. Generally, we use a set of websites as training data and learn a prediction model to classify a website as malicious or benign. To fully harness the power of data-driven approaches, it is essential to comprehend and improve various stages of the malicious website detection pipeline, including data preprocessing, feature extraction, model construction and technology extension.

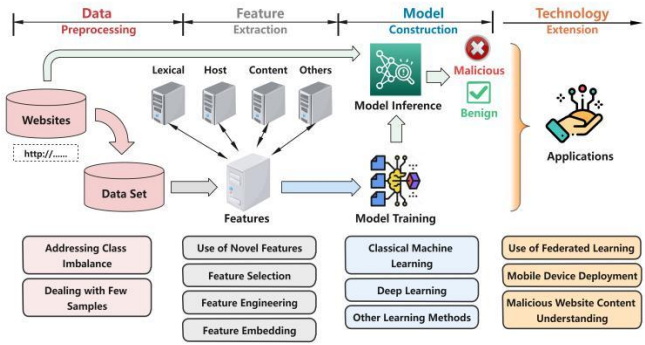

Fig. 3. Data-feature-model-extension: the malicious website detection pipeline of data-driven methods.

Figure 3 illustrates the different components of the pipeline of data-driven methods. Specifically, we will discuss how to address class imbalance and deal with few samples in data preprocessing. We will also cover the use of novel features, feature selection, feature engineering, and feature embedding in feature extraction. In addition, we will examine traditional machine learning, deep learning, and other learning techniques for model training. Furthermore, we will present recent technology extension for preventing malicious websites, such as federated learning, mobile devices, and content understanding.

By summarizing the latest research developments in data-driven approaches for malicious website detection, this paper aims to provide a comprehensive overview of the state-of-the-art in this field.

### A. Improvements in Data Preprocessing Techniques

Data preprocessing is the cornerstone of a data-driven approach. In malicious website detection, current research focuses on solving class imbalance and dealing with few-sample problems to prepare the data for subsequent modeling.

#### 1) Addressing Class Imbalance

Class imbalance refers to the situation where the number of samples in one class is significantly smaller than that in the other classes. In malicious website detection, the number of malicious websites is often much smaller than that of benign websites. Class imbalance can cause the machine learning algorithms to be biased towards the majority class, resulting in poor performance in detecting the minority class.

Several techniques can be used to address class imbalance, such as over-sampling the minority class, down-sampling the majority class, and generating synthetic samples. One popular technique is the Synthetic Minority Over-sampling Technique (SMOTE), which generates synthetic samples by interpolating between neighboring samples of the minority class [5].

According to the experiments presented in [6], after balancing the data, the random forest algorithm using the over-sampling technique showed the best results in all evaluation metrics for the benign and malicious website feature dataset.

#### 2) Dealing with Few Samples

Although many companies have web filtering algorithms that flag certain URLs as malicious, they will not allow other parties access to their cybersecurity data due to privacy concerns. Therefore, in malicious website detection, collecting a large number of high-quality samples is often a challenging task. Few samples can lead to overfitting of machine learning algorithms, resulting in poor generalization performance. We typically use data augmentation techniques to address the issue of few samples in malicious website detection.

*a) ML-based methods:* Vecile et al. [7] proposes using a machine learning method to generate new synthetic URLs that are indistinguishable from the data they replace, which is achieved by training two character-level LSTM models, one for generating malicious URLs and another for generating benign URLs.

*b) GAN-based methods:* Recently, GAN-based models have been attractive in data augmentation for malicious URLs. The overall motivation is to use Generator network (G) to generate fake URL samples and Discriminator network (D) to tell the difference between the real and the fake URL samples. In the naive GAN approach, Both D and G are parameterized by a neural network and trained by the loss function [8]:

$$\underset{G}{\text{Min}}\,\underset{D}{\text{max}}\,L(D,G) = \mathbb{E}_x[logD(x)] + \mathbb{E}_z[\log(1 - D(G(z)))] \quad (1)$$

In order to adapt the method for the application of malicious website detection, Pham et al. [9] propose to train a GAN named WGAN-GP for generating malicious URLs from the available URL data set. The WGAN-GP loss function is set based on the original critic loss of Wasserstein GAN (The Wasserstein distance) combined with a proportion of gradient penalty:

$$L = \mathbb{E}_{\hat{v} \sim \mathbb{P}_g}[D(\hat{v})] - \mathbb{E}_{v \sim \mathbb{P}_r}[D(v)] + \lambda \mathbb{E}_{\hat{v} \sim \mathbb{P}_{\hat{v}}}[(\|\nabla_{\hat{v}}\|_2 - 1)^2] \quad (2)$$

### B. Improvements in Feature Extraction Techniques

Feature extraction is a crucial step in machine learning algorithms for malicious website detection, which utilizes a combination of business experience and intelligent technology. It involves selecting or constructing features that can best represent the underlying multidimensional characteristics of the data, with the goal of enhancing the efficiency and the accuracy of the machine learning algorithms. Recent studies on website features have focused on the use of novel features, feature selection, feature engineering and feature embedding techniques.

*1) Use of Novel Features*

The effectiveness of feature extraction depends essentially on the features used. For malicious website detection, researchers have proposed several types of features, including URL-based features (U), host-based features (H), content-based features, and others (O) [10]. Due to the characteristics of machine learning or deep learning algorithms, content-based features can be further divided into visual content features (V) and textual content features (T). Table I lists commonly used website feature sources by category.

TABLE I. COMMONLY USED WEBSITE FEATURES

| Feature Type | Typical Feature Sources |
|---|---|
| URL-based (*U*) | URL string, N-gram features, word-level features, character-level features, entropy, etc. |
| Host-based (*H*) | IP address properties, WHOIS, location, DNS, domain properties, etc. |
| Visual Content (*V*) | Images, screenshots, logos, etc. |
| Textual Content (*T*) | HTML, DOM, JavaScript, CSS, titles, keywords, etc. |
| Others (*O*) | Alexa, PageRank, etc. |

In addition to the commonly used website features mentioned above, researchers have explored the use of novel features in recent years to improve the accuracy of detection. These features can capture the unique characteristics of malicious websites and help to incorporate the latest prior knowledge and domain expertise into machine learning models.

In the study of [11], hyperlink indicators along with URL-based features are used to construct the detection model. In another study, Chen et al. [12] demonstrate that adding four novel features—including Communication Countries Rank (CCR)—can improve the performance and robustness of the machine learning model. Wang et al. [13] first propose to use the text information in the image by OCR along with the image itself, which often has obvious semantic features and clearly points to gambling websites. Rozi et al. [14] propose a framework named AST-JS to enrich the webpage features for improved detection models using a JavaScript representation based on the abstract syntax tree. In addition, Liu et al. [15] propose a comprehensive and interpretable CASE feature framework with complete consideration of the social engineering principles.

*2) Feature Selection*

Feature selection is a crucial process in malicious website detection that involves selecting a subset of relevant features from the original feature set. This process can help to remove irrelevant or redundant features, which can improve the efficiency and accuracy of the detection model.

Several techniques have been commonly used for feature selection in malicious website detection, such as chi-square, Information Gain (IG), correlation feature selection (CFS) and wrapper method. Mohanty et al. [16] proposed a method called MFBFST that uses six feature selection techniques and improves the accuracy of malicious website detection in various widely used machine learning algorithms.

Moreover, researchers have explored the use of evolutionary algorithms, such as genetic algorithms and particle swarm optimization. For example, Khurma et al. [17] propose a new phishing detection system based on the Salp Swarm Algorithm (SSA). According to their experimental results, the binary SSA with X-TFs outperforms previous approaches and achieves the highest accuracy for identifying phishing websites.

*3) Feature Engineering*

The research history of data-driven methods indicates that a simple concatenation of different features may not be sufficient to achieve satisfactory results, since different types of features often belong to distinct feature spaces and vary significantly from one another. Feature engineering involves transforming the original features into a new unified feature space that is more suitable for modeling, thereby improving the performance of machine learning algorithms. Commonly used techniques include feature encoding, feature construction, and feature dimensionality reduction, such as principal component analysis (PCA), linear discriminant analysis (LDA), and kernel methods.

Recently, novel feature engineering approaches have been proposed for malicious website detection. Li et al. [18] propose a combination of linear and non-linear feature space transformation methods, which is achieved through a two-stage distance metric learning approach and introducing Nyström method for kernel approximation.

### 4) Feature Embedding

Feature embedding can be considered as an advanced version of feature engineering, which involves representing the original features as dense vectors in a low-dimensional space to improve the performance of the detection model. Several studies have explored the use of feature embedding in malicious website detection.

For URL representation, some researchers have proposed using self-supervised learning to obtain URL embedding vectors. In the proposed framework by [19], the inherent feature vector representations of a raw URL are automatically extracted by Variational Autoencoders (VAE) through reconstructing the original input URL to enhance phishing URL detection.

For content representation, Chen et al. [20] use the Doc2Vec model to obtain textual embeddings and use the local spatial improved bag-of-visual-words (Spa-BoVW) model to obtain visual embeddings of websites.

In addition, Chaiban et al. [21] observe and evaluate the combinations of normal features and feature embeddings involving image embeddings, content embeddings and URL embeddings. These feature embeddings are extracted using MobileNetV2 [22], CodeBERT [23, 24], DistilBERT [25] and Longformer [26]. The experimental results indicate that URL embeddings are the most significant feature for improving malicious website detection models; the gain of embedding-related features is above 1% compared to commonly used content-related features in terms of model performance.

## C. Improvements in Model Construction

With the rapid development of the AI era, the methods for model construction have also been significantly enriched. We mainly focus on recent improvements in machine learning model construction for malicious website detection and the latest research in this field.

### 1) Traditional Machine Learning

Traditional machine learning algorithms include decision trees (DTs), logistic regression (LR), support vector machines (SVMs), random forests (RFs), and Naive Bayes (NBs). These algorithms are relatively easy to understand and interpret, and they can be effective for smaller datasets. However, they may not be able to capture the complex patterns in the data, and their performance may degrade for larger datasets.

Researchers often use ensemble learning models, such as Random Forests (RFs), XGBoost [27], and LightGBM [28], to address the limitations of traditional methods. Additionally, [29-32] respectively propose their ensemble learning frameworks for malicious website detection using multiple heterogeneous machine learning models.

### 2) Deep Learning

Deep learning is a subfield of machine learning that involves training artificial neural networks with multiple layers. It has become a mainstream technology due to its superior ability to learn a hierarchy of representations from low-level to high-level features, which are gradually combined to form more abstract and meaningful representations of the input data [33].

In malicious website detection, in addition to commonly used convolutional neural networks (CNNs), recurrent neural networks (RNNs), gated recurrent units (GRUs) and long short-term memory networks (LSTMs), recent research has explored the use of graph neural networks (GNNs) [34-36], pre-trained language models (PLMs) and multimodal learning, as shown in Table II. Research by [37-40] propose to construct Bidirectional Encoder Representation from Transformers (BERT) or its improved version for malicious website detection. [13, 38] propose using advanced attention mechanisms to integrate two of the three modalities for websites (text, image and code), which could jointly represent the information such that the detection model is able to capture the correlation structure between different modalities.

TABLE II. DEEP LEARNING MODELS IN MALICIOUS WEBSITE DETECTION RESEARCH OVER THE PAST THREE YEARS

| Methods | U | H | T | V | O | Models |
|---|---|---|---|---|---|---|
| Texception (2020) [41] | √ | | | | | CNN |
| Web2Vec (2020) [42] | √ | | √ | | | CNN, LSTM |
| URL-Visual-CNN (2020) [43] | √ | | | √ | | CNN |
| IndRNN-CapsNet (2021) [44] | √ | | | √ | | RNN, CapsNet |
| HTML-GNN (2021) [35] | | | √ | | | GNN |
| PhishGNN (2021) [34] | √ | √ | √ | | √ | GNN |
| BERT/ELECTRA (2021) [37] | √ | | | | | PLM |
| URLTran (2021) [40] | √ | | | | | PLM |
| MDF/MFF/MIF (2022) [45] | √ | | √ | √ | | CNN |
| MHSA-BiLSTM-MCTC (2022) [46] | √ | √ | √ | | | CNN, LSTM |
| DA-BiGRU (2022) [47] | √ | | | | | GRU |
| PhishDet (2022) [36] | √ | | √ | | | CNN, LSTM, GNN |
| P-BERT (2022) [39] | √ | | | | | PLM |
| ResNet34-BiLSTM (2022) [13] | | | √ | √ | | Multimodal Learning |
| MM-ConvBERT-LMS (2023) [38] | √ | | √ | | | Multimodal Learning, PLM |
| DeepBF (2023) [48] | √ | | | | | Learned Bloom Filter, CNN |

### 3) Other Learning Methods

In addition to traditional machine learning and deep learning, other learning methods have also been explored in the field of malicious website detection, such as similarity learning, semi-supervised learning and reinforcement learning.

*a) Similarity learning:* The objective of similarity learning is to learn how similar two instances are, or in the context of our study, how similar two websites are. According to this approach, a set of protected URLs is established along with a set of suspicious URLs that may attempt to imitate the protected URLs. The method involves developing a measure of similarity, extracting webpage features (including textual features or visual features) from the websites to be detected, and then calculating the similarity between suspicions URLs with the protected URLs. If the similarity exceeds a certain threshold, the website is identified as malicious [10]. For textual similarity learning, Mao et al. [49] obtain and compute vector-based features from website CSS codes and detected phishing websites based on textual vector similarity. Tanaka et al. [50] propose a simpler method of extracting website features solely from the type and size of network resources contained in the web access log data. Purwanto et al. [51] propose the PhishSim framework and a non-parametric similarity measure, NCD, for detecting phishing websites based on the normalized compression distance of HTML codes without feature extraction. However, some criminals may present visually similar web pages with vastly different code scripts, which poses technical challenges in inferring web page similarity and affects the detection accuracy. For visual similarity learning, Abdelnabi et al. [52] propose VisualPhishNet, which utilizes triplet network and convolutional neural sub-networks to learn similarity between snapshots of the same website and dissimilarity between snapshots of different websites, and iteratively find hard samples for training based on the latest checkpoint of the model. Trinh et al. [53] address the complexity of the visual feature extraction process for web pages and proposed transfer learning techniques based on pre-trained image classification models for phishing website detection.

*b) Semi-supervised learning:* The objective of semi-supervised learning is to use both labeled and unlabeled data to train a machine learning model. It has been used in malicious website detection to leverage the large amount of unlabeled website samples available, which can improve the performance of the detection model and reduce the demand for manually labeled data. Zhang et al. [54] propose a PU learning (Positive and Unlabeled learning) based system for malicious URL detection by combining two-stage strategy and cost-sensitive strategy.

*c) Reinforcement learning:* The objective of reinforcement learning involves an agent interacting with an environment to learn how to take actions that maximize a reward signal. Chatterjee et al. [55] propose to model the detection of phishing websites through deep Q network (DQN), The agent learns the following reward function from the given input website to perform classification tasks:

$$R_{c=} \sum_{k=1}^{\infty} \gamma^k \cdot r_{t+k} \quad (3)$$

D. *Recent Technology Extension*

In addition to classical data-driven techniques mentioned above, recent research has focused on developing novel approaches or applications for detecting and preventing malicious websites. In this section, we highlight three recent advances in technology expansion for malicious website detection.

*1) Use of Federated Learning*
Federated learning is a decentralized machine learning approach that allows multiple devices to collaboratively train a model without sharing their data. For example, FedAvg [56] is a commonly federated learning algorithm, which allows data collaborators to update the model using a local random gradient descent algorithm, and then aggregate parameters at the central node . Its loss function is as follows:

$$f(w^*) = min\{f(w): = \frac{1}{M}\sum_{n=1}^{M} E[f(w:x, x \in n)]\} \quad (4)$$

This technology has been proposed for malicious website detection to address the challenges of collecting and sharing large amounts of sensitive data. By using federated learning, multiple devices can train a model on their local data and contribute to a global model without sharing their data, which can improve the accuracy and efficiency of malicious website detection while maintaining data privacy and security.

Ongun et al. [57] propose a federated machine learning framework for global threat detection called CELEST, which uses federated learning techniques for generating feature embeddings by federated FastText and training a neural network for detection of malicious URLs. Makkar et al. [58] first propose to investigate the applicability of Internet attack detection through federated machine learning. The experimental results show that the application of federated learning to satisfy customer search queries by detecting malicious spam images suits practical scenarios.

*2) Mobile Device Deployment*
As the use of mobile phones continues to grow, these devices are being used rapidly for accessing the web and many online services. However, the security mechanisms available in smartphones are not yet fully developed. To match the capabilities of the mobile devices, the browsers on them are extremely simple, and their security features have also been scaled back. Therefore, detection of the malicious website is utterly crucial and different from the previously existing techniques, which are used on desktop systems and servers. One recent approach is to deploy detection models directly on mobile devices, which can provide faster and more efficient detection without relying on network connectivity.

Wei et al. [59] propose a deep neural network with a lightweight architecture to detect malicious URLs, which enables real-time and energy-efficient detection. Jain et al. [60] propose a mobile-based efficient malicious website detection system named APuML through various stages, including data feed, DNS and machine learning. The authors demonstrate that the system achieve high detection accuracy with low response time. Haynes et al. [37] apply deep transformer (BERT and ELECTRA) for phishing detection which is suitable for running on mobile devices and is effortlessly deployable for real-time detection. In the study of Liu et al. [61], a multidevice load optimization approach and an automatic deep feature extraction approach are presented to improve the efficiency and accuracy, with flexible deployment on edge nodes and servers for resource optimization and real-time detection.

*3) Malicious Website Content Understanding*
In practical public security scenarios, after completing the classification task on a massive number of websites, a batch of suspicious websites can be selected. However, at this point, only a single-dimensional classification label result (malicious or benign) about the website can be obtained. Therefore, it is often necessary to further understand and explore every part of the website content to supplement the interpretability and practicality of the model, which can provide strong support for subsequent business operations such as clues mining, investigation and evidence collection. Unlike ordinary documents, in the business of understanding the content of malicious websites, there are more complex modal media contents such as images, text, and tables, which pose great challenges to data-driven methods. Currently, research mainly focuses on the general understanding of webpage content or content understanding in other fields, and there is little research on the content understanding of malicious websites.

Lin et al. [62] proposed a hybrid deep learning system, Phishpedia, which uses target detection models and transfer learning-based Siamese networks to perform logo recognition and brand variant matching. It can achieve good performance and interpretability under zero-shot conditions and complete the analysis of phishing webpage snapshots and their URLs within 0.2 seconds. Gu et al. [63] consider that the fraudulent website often includes complex and rich page hierarchy structures, such as registration columns, login columns, transfer columns, customer service chat boxes, etc. Therefore, they propose the multi-modal content understanding model named XYLayoutLM, which improves the error understanding caused by the complex structure and long text in the automatic browsing process, and thus improves the performance of various sub-businesses in the anti-fraud business, such as website parsing and risk qualitative analysis.

## III. CHALLENGES AND FUTURE DIRECTIONS

### A. Limitations and Challenges

Although there have been significant advancements in the field of data-driven approaches for malicious website detection, there are still some limitations and challenges that need to be addressed.

*1) The need of high-quality datasets*

*a) Scale:* The dataset should be large enough to encompass a broad range of malicious websites. However, malicious websites tend to expire quickly, and obtaining a real-time dataset of active malicious websites is challenging due to privacy policy restrictions.

*b) Annotation:* The annotation of the dataset requires significant human effort and time. Moreover, the expertise level of the annotators must be high, as there exist gray areas between different types of websites that are challenging to determine. Unfortunately, semi-supervised and weakly supervised methods have not yet been maturely applied in the field of malicious website detection.

*2) The utilization of heterogeneous features*

*a) Feature fusion:* Considering that website data has multiple complex attributes, research on how to unify and represent heterogeneous, multi-modal, and multi-source features remains a challenging task, which is the foundation for obtaining better data-driven models.

*b) Scenario adaption:* Different business scenarios have different requirements for feature extraction of malicious websites. There is still little work on selecting, processing, and representing heterogeneous features that are adapted to scenarios, based on understanding the different downstream tasks.

*c) Efficiency:* For the collection of massive website content features, including text and images, significant computational resources and time are required, which can affect the online processing capabilities of malicious website detection.

*3) Concerns about models with trustworthy performance*

*a) Generalization:* Since malicious websites update and evolve rapidly, detection models may perform well on known types of samples, but may struggle to detect new and previously unseen malicious information. Developing models that can quickly adapt to new criminal methods is important for staying ahead of criminal.

*b) Privacy security:* Some detection techniques may rely on collecting and analyzing user data, raising concerns about privacy and data security. Finding ways to balance effective detection with protecting user privacy is an ongoing challenge.

*c) Interpretability:* Interpretability is crucial in building trust in the models and enabling their adoption in practice. However, many deep learning models used for malicious website detection lack interpretability and are considered black-box models, making it challenging to understand their decision-making process. The lack of interpretability can hinder the deployment of the model in real-world scenarios, where transparency and accountability are essential.

### B. Future Directions and Potential Solutions

To address the challenges of current techniques, future research can focus on developing novel methods or the combination application of existing methods that can handle the above-mentioned issues.

For the need of high-quality datasets, one future direction is the development of semi-supervised and weakly supervised techniques that can reduce the dependence on labeled data. Another direction is the use of pre-trained self-supervised techniques to make full use of massive unlabeled data on the Internet.

For the utilization of heterogeneous features, one potential solution is to explore advanced methods for feature fusion of multi-modal data and develop a theoretical framework for unified feature extraction. Additionally, to develop methods for real-time and continuous lightweight feature extraction, research can focus on distributed and parallel computing techniques to accelerate the content feature collection.

For concerns about models with trustworthy performance, the integration of human expert knowledge and heuristics into the machine leaning detection models can improve their performance and interpretability. Another solution is to build fine-grained classification tasks instead of binary classification tasks of malicious or benign websites. Furthermore, researchers can explore the feasibility of federated learning and blockchain technology to provide a secure and decentralized platform for data sharing among researchers and organizations without compromising privacy.

In the future, we can further focus on technology extension of data-driven methods in multiple dimensions, such as: deploying device-cloud collaborative models to achieve real-time monitoring of malicious websites; providing intelligent clues for preventing telecom fraud, illegal fundraising, online gambling and other criminal activities. The data-driven approaches in malicious website detection will continue to expand into novel application fields.

IV. CONCLUSION

In conclusion, data-driven approaches have emerged as a promising solution to the challenges and limitations of traditional techniques for malicious website detection. This review is developed around the theme of data-driven approaches, focusing on the data-feature-model-extension pipeline. We discussed various recent improvements in data preprocessing, feature extraction, and model construction. Additionally, we also highlighted recent technology extension. Although there have been significant improvements in recent years, there are still several challenges that need to be addressed, such as the need of high-quality datasets, the utilization of heterogeneous features and concerns about models with trustworthy performance. Therefore, future research can focus on developing novel methods or the combination application of existing methods to handle the issues. Given a comprehensive data-driven malicious website detection architecture, followed by various improvements for each phase discussed in this paper, we believe that it is appropriate for researchers to optimize data processing and feature extraction methods, as well as building trustworthy models with a wide range of extended technical applications. Considering the ongoing in-depth research in this field, more practical data-driven approaches will contribute significantly to creating a safer online environment for users and enterprises.

ACKNOWLEDGMENT

This work is funded by National Key Research and Development Project (Grant No: 2022YFB2703104)..